\documentclass[a4paper,11pt]{article}
\usepackage{pos}

\usepackage{natbib}
\title{Diquarks in lattice QCD}

\author*[a]{Anthony Francis}

\affiliation[a]{National Yang Ming Chiao Tung University,\\
  1001 Daxue Road, Hsinchu, Taiwan}

\emailAdd{afrancis@nycu.edu.tw}

\abstract{Diquarks are often invoked as QCD effective degrees of freedom to describe baryons as well as certain exotic hadrons in phenomenology. However, even though they are successful in describing many of these low lying QCD states, they and their properties have been difficult to pin down. Here we present progress in studying diquarks in a gauge-invariant setup through embedding them in a parent hadron containing a heavy spectator quark using lattice QCD calculations.}

\FullConference{The XVIth Quark Confinement and the Hadron Spectrum Conference (QCHSC24)\\
 19-24 August, 2024\\
 Cairns Convention Centre, Cairns, Queensland, Australia\\}

\begin{document}
\maketitle

\section{Motivation}

Diquarks as effective degrees of freedom in hadrons have long been used to explain and describe the masses of low-lying baryons \cite{Lichtenberg:1967zz}. Their success in correctly reproducing them has been a compelling argument to consider diquarks and their properties as key features of QCD \cite{Jaffe:2004ph}. In recent times they have also played a role in motivating certain exotic states and especially doubly heavy tetraquarks \cite{Bicudo:2022cqi,Francis:2024fwf}.
However, providing evidence for their existence in experiment and determining their properties in theory remains difficult due to their suggested nature as colored QCD degrees of freedom. In the following we present our recent and ongoing work to approach diquarks through embedding and isolating them in parent hadrons with spectator quarks. The work presented represents an extension of our work in \cite{Francis:2021vrr}.

\section{Probes of diquark properties}

Formally the interpolating operator of a diquark may be written as
\begin{equation} 
    D_\Gamma =q^c C \Gamma q'
\end{equation}
where $q,q'$ denote different quark flavors, $c,C$ denote charge conjugation and $\Gamma$ acts on Dirac space. Choosing different $\Gamma$ combinations in the interpolating operator allows the definition of their effective quantum numbers as:
\begin{center}
\begin{tabular}{lllll}
       $J^P$ & color & flavor & $\Gamma$ & label\\ \hline
         {$0^+$}& {$\bar 3$}&{$\bar 3$}& {$\gamma_5$, $\gamma_0\gamma_5$} & "good"\\
         {$1^+$}& {$\bar 3$}&{$6$}& {$\gamma_i$, $\sigma_{i0}$} & "bad"\\ 
         $0^-$& $\bar 3$&$6$& $1\!\!1$, $\gamma_0$ &  "not-even-bad"\\ 
         $1^-$& $\bar 3$&$\bar 3$& $\gamma_i\gamma_5$, $\sigma_{ij}$ & "not-even-bad"
\end{tabular}
\end{center}
Of these different diquark configurations the $J^P=0^{+};~\overline{{3}}_{\mathrm{c}};~\overline{{3}}_{\mathrm{f}}$ is especially interesting since its flavor color quantum numbers could allow for an attraction between the constituent quarks $q,q'$ that is not accessible in the other configurations. For this reason this diquark configurations is also referred to as the "good" diquark while the others are labeled "bad". For the $J^P=0^-,1^-$ cases also the terminology "not-even-bad" is used to distinguish them further.
Since quarks in good diquarks are predicted to attract each other this implies the observation of a large mass splitting between bad/good and not-even-bad/good channels. Indeed, these splittings would be physical properties of QCD \cite{Jaffe:2004ph}.
They should be dependent on the masses of the constituent quarks. In the case of $q=u,q'=d$ we would expect a stronger attraction in the good diquark than in $q=u,q'=s$. Furthermore, in the limit $q\rightarrow Q$ where $m_Q=\infty$ considerations based on heavy-quark spin-symmetry (HQSS) suggest that when the heavy quark spins decouple a diquark acts as an antiquark $[QQ]\leftrightarrow \bar Q'$. 



\textbf{Mass differences as gauge-invariant probes.} A problem for searching for diquarks and confirming the above pictures is their colored nature. Since they cannot be created in isolation in experiments the good diquark attraction can only be studied indirectly through hadrons that they might be embedded in.
This is also a problem in principle for lattice QCD since only gauge-invariant quantities are directly calculable in correlation functions. One could in principle fix a gauge \cite{Hess:1998sd,Bi:2015ifa,Babich:2007ah,Teo:1992zu,Alexandrou:2002nn}, however, then the determined diquark properties are gauge dependent and need to be further verified.
Here, we propose to use a different approach in which we use the capability of lattice calculations to vary quark masses at will and also to fix $m_Q=\infty$ by choosing a static quark. Then we may access mass differences of the type alluded to above from first principles by embedding the diquark in static-light-light or static-light-strange baryons \cite{Alexandrou:2005zn,Alexandrou:2006cq,Orginos:2005vr,Green:2010vc,Francis:2021vrr}. Considering also the static-light meson case, their correlation functions generically read:
\begin{align}
C_{\Gamma}(t)&=\sum_{\vec x} \Big\langle [D_\Gamma Q](\vec x,t)~[D_\Gamma Q]^\dagger(\vec 0,0)  \Big\rangle~~~~=\text{static-light-light baryon}\\
C^M_{\Gamma}(t)&=\sum_{\vec x} \Big\langle [\bar Q \Gamma q](\vec x,t)~[\bar Q \Gamma q]^\dagger(\vec 0,0)  \Big\rangle~~~~=\text{static-light meson}
\end{align}
In the case of $Q$ being a static quark we may exploit an effective mass decomposition and write the long-time asymptotic region of the correlator as:
\begin{equation}
C_{\Gamma}(t)\sim \exp\left[-t\left(m_{D_{\Gamma}} + m_Q + \mathcal{O}(m_Q^{-1})\right)\right]~,~~~C^M_{\Gamma}(t)\sim \exp\left[-t\left(m_{q} + m_Q + \mathcal{O}(m_Q^{-1})\right)\right]
\label{eq:massdecomp}
\end{equation}
This poses an opportunity to make gauge-invariant observations for diquarks because choosing a static spectator quark, i.e. $m_Q=\infty$, implies $\mathcal{O}(m_Q^{-1})=0$ and it can be canceled out exactly in  appropriate differences.
For different light quark flavor and diquark combinations we can thus access the mass differences that we are interested in directly:
\begin{equation}
m_{\Gamma}^{qq'Q}(t) - m_{\gamma_5}^{qq'Q}(t) ~~ \stackrel{t\rightarrow \infty}{\longrightarrow} ~~ M^{qq'}_{bad} - M^{qq'}_{good}
\end{equation}
where bad implies any of the three bad diquark operators, $m(t)$ is the effective mass of the correlator and $M$ is the long-time asymptotic value extracted from it. One can go further and also define static-light-light baryon and static-light meson differences as:
\begin{equation}
m_{\Gamma=\gamma_5}^{qq'Q}(t) - m_{\gamma_5}^{q'\bar{Q}}(t) ~~ \stackrel{t\rightarrow \infty}{\longrightarrow} ~~ M^{qq'}_{good} - M^{q}
\end{equation}
The argument rests on the applicability of the mass decomposition in Eq.~\ref{eq:massdecomp} and clearly shows its limitations: 
While the $qq'Q$ case poses a clear diquark probe, other embedded baryons, such as $qQQ'$ or $qq'q''$ are less clear. In the former HQSS relates it to a static-light meson, and in the latter we can expect (strong) interaction effects from the $\mathcal{O}(m_Q^{-1})$ term. From the point of view of wanting to study whether diquarks are also formed in baryons such as $\Lambda_{s,c}$ or the nucleon, this approach is stretched. In the case of studying diquarks in isolation by embedding it with a static quark the approach is more justified.

Apart from testing diquark properties, embeddings with a static spectator can also be used for other purposes. In the following we explore one to test the large quark mass regime. This work is still preliminary.
The aim is to quantify how well the quarks do in fact decouple and hadrons or other effective degrees of freedom become degenerate in this limit.
To this extent we first consider the mass difference between the $qq'Q$ baryon and the $q\bar{Q}$ meson 
\begin{equation}
\delta_{q}(t)={C_{\gamma_{i}}^{qqQ}(t) - C_{\gamma_{i}}^{q\bar{Q}}(t)}~~\rightsquigarrow \textrm{ determine on the lattice: } M_{\delta_{q}}
\end{equation}
where the quarks can be $q\in l,s,c,b$, and the diquark should have the same quark flavors. The goal is to determine the mass splitting $m_{\delta_{q}}$ from the long-time asymptotic behavior. In addition we determine the $\Omega_q$ baryon mass using standard methods.
Relating these two quantities leads one to consider: 
\begin{equation}
\frac{M_{\Omega_{q}} }{ M_{\delta_{q}}} = \frac{ M^{qqq} }{ M^{qqQ}-M^{q\bar Q}} ~~{\longrightarrow}~~~\text{HQSS: }~~ \frac{M_{\Omega_{q}}}{ M_{\delta_{q}} } = 3
\end{equation}
In the HQSS limit the expected degeneracy pattern of the (dressed) quarks would imply this ratio to go to the value 3, as the denominator gives an estimator of the diquark-quark mass difference. 

As part of testing this aspect of HQSS this also tests the, some times used, technical assumption in lattice QCD that heavy hadrons can be decomposed into dressed quarks by dividing their number. This happens, for example, in calculations using (lattice) NRQCD in certain cases. Here the mass parameters of the calculation are tuned to reproduce the kinetic masses of the target quarkonia. This is usually done by extracting the fitted coefficients of the momentum-dispersion relation in $M_{Q\bar Q}(\vec p)$, where the kinetic mass drives the slope in a linear fit to the $\vec p^2$ dependence and the rest mass is the $\vec p=0$ intercept. The latter value is unphysical in the sense that it does not directly relate to the target masses. In quarkonia this is not an issue since one can relate directly to the tuning. Also in mass differences of hadrons this will drop out. The problem arises when studying observables where this factor remains, either uncanceled or through an imbalance of the number tuned and to be studied heavy quarks in the target hadron. In these cases a commonly used trick is to take the rest mass parameter and divide it such that one can estimate its contribution for one heavy quark. The alternative would be to tune determine the kinetic mass from the momentum-dispersion relation, however, this can be forbidding due to noise and cost issues.
With the suggested test we can check in which mass regimes this is better and worse justified. 

\begin{figure}
\centering
\includegraphics[width=0.4\textwidth]{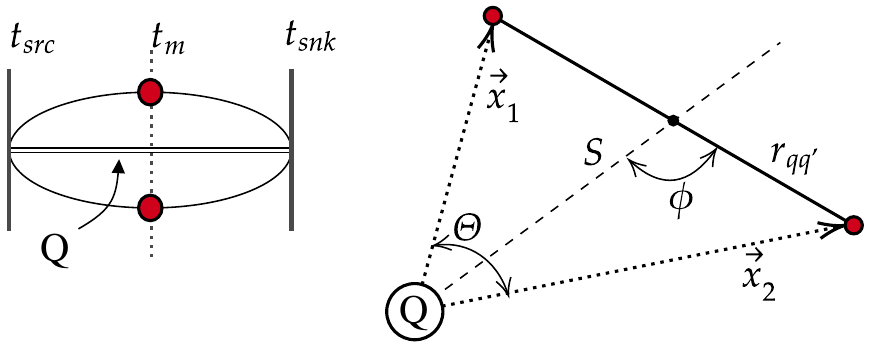}\hspace{10ex}
\includegraphics[width=0.4\textwidth]{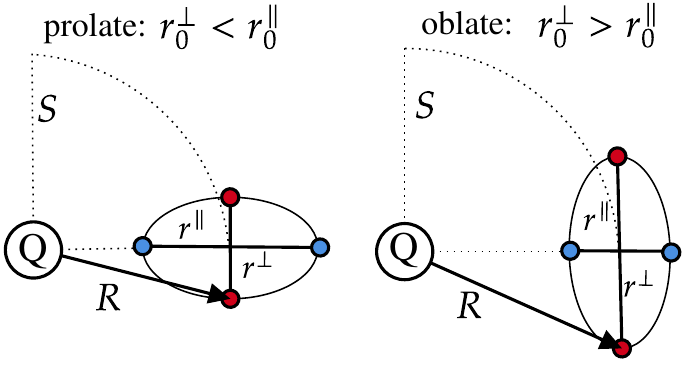}
\caption{Left: Geometry of the density-density correlations used to study the diquark attractive effect. We define $\vec r_{ud}=\vec x_2 - \vec x_1$ and $\vec S=(\vec x_1 - \vec x_2)/2$. Further explanations are given in the text. Right: When $|\vec x_1|\neq|\vec x_2|$, we can introduce a new vector $\vec R$ that points to the radial/tangential deviation from $\vec S$. This allows to probe the shape of the diquark correlations. }
\label{fig:geom1}
\end{figure}


\textbf{Density correlations for the good diquark attraction.} While studying the mass differences already gives indications on the effectiveness of the good-bad diquark picture, they do not provide a direct
handle on whether there is a uniquely effective attraction between the two quarks in a good diquark over all other configurations.
Further insights in this direction can be obtained from lattice QCD calculations by studying the diquarks in the static-light-light embedding through density-density correlations \cite{Alexandrou:2005zn,Francis:2021vrr}:
\begin{align}
   { C_{\Gamma}^{dd}(\vec x_1, \vec x_2, t)=
    \Big\langle D_\Gamma(\vec 0,2t)~
    \rho(\vec x_1,t)\rho(\vec x_2,t)~
    D_\Gamma^\dagger(\vec 0,0)  \Big\rangle}
\end{align}
where $\rho(\vec x,t)=\bar{q}(\vec x,t)\gamma_0 q(\vec x,t)$, $t_m=(t_{snk}+t_{src})/2$ and $t_m$ is chosen to minimize excited state contamination as much as possible. 
By studying the spatial correlations of these two current insertions around the spectator we can get a picture of the interactions happening within the baryon. As such we expect these correlations to behave as exponential decays whose coefficients define effective correlation lengths, or effective radii, $r_0$.
In the static-light-light case, we in particular expect no further perturbation through interactions with the spectator and that there is no impact on the correlation direction that we choose, either radially towards the spectator or tangentially to it. 

The geometric configurations of the correlations between the relative positions of the two light quarks to the static quark that we study are sketched in Fig.~\ref{fig:geom1} (left). At this point we introduce the notation
    ${\rho_2(r_{ud},S,\phi; \Gamma) = C_{\Gamma}^{dd}(\vec x_1, \vec x_2, t_m)}.$
An interesting case is the situation when $|\vec x_{1}|\neq |\vec x_{2}|$, as sketched in Fig.~\ref{fig:geom1} (right). Introducing $\vec R=\vec S + \vec r(\phi)$, we study the two cases:
(1.) $\phi=\pi$, radial correlation. Where we expect the correlation to go as $\sim\exp[ (r_0^\parallel)^{-1} t]$.  (2.)
$\phi=\pi/2$, tangential correlation. Where we expect the correlation to go as $\sim\exp[ (r_0^\perp)^{-1} t]$.
Measuring both $r_0^\perp$ and $r_0^\parallel$ gives information between the quarks in the embedded hadron. As such, a visible exponential decay indicates attraction between the quarks in the diquarks. A match between the radial and tangential radii ($r_0^\perp/r_0^\parallel = 1$) indicates a spherical shape of the diquarks with no further polarization through the spectator. At the same time a mismatch $r_0^\perp/r_0^\parallel \neq 1$ indicates a prolate or oblate shape and there for significant polarization and interactions with the spectator.

\begin{figure}
\centering
\includegraphics[width=0.46\textwidth]{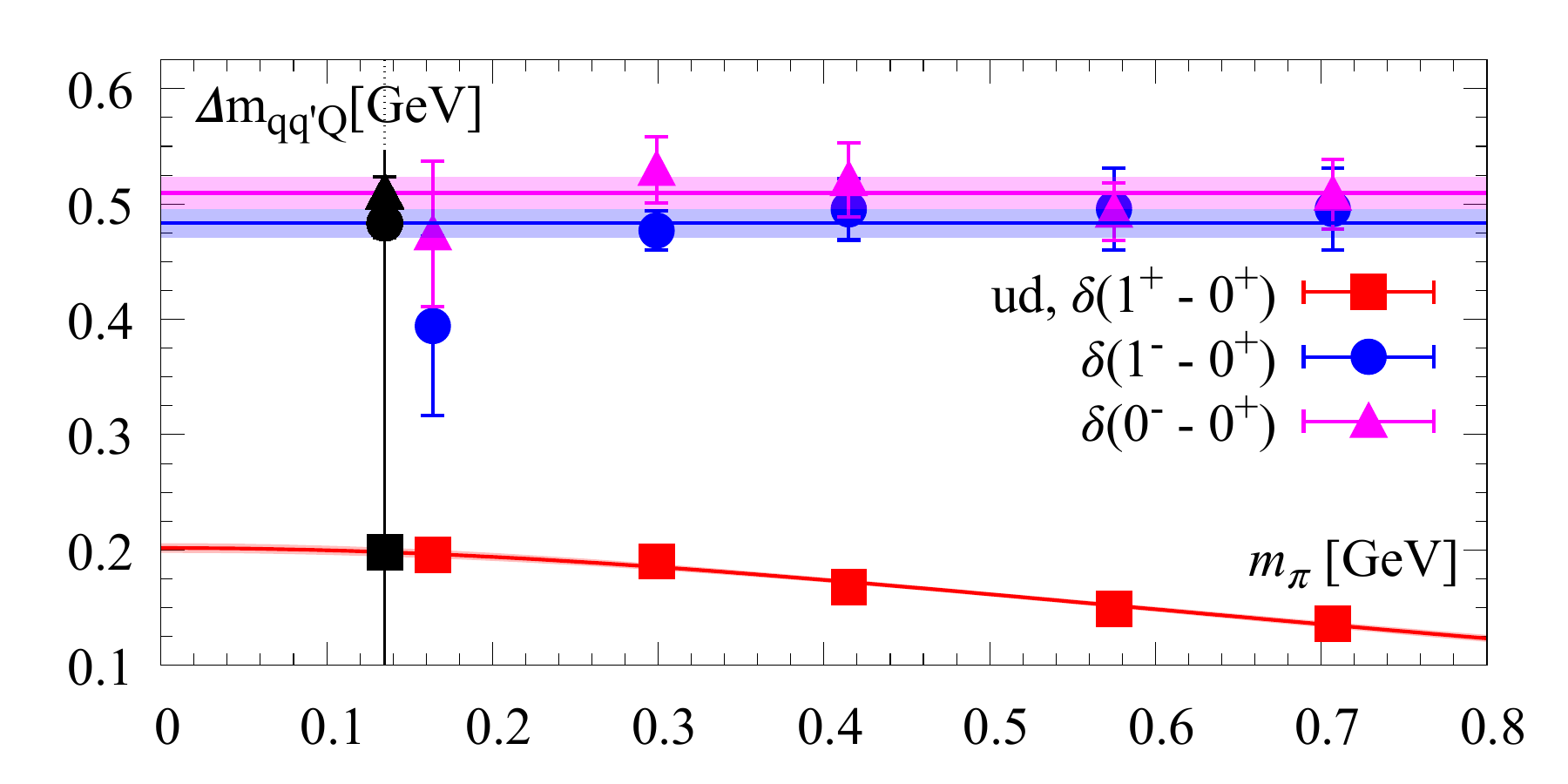}
\includegraphics[width=0.46\textwidth]{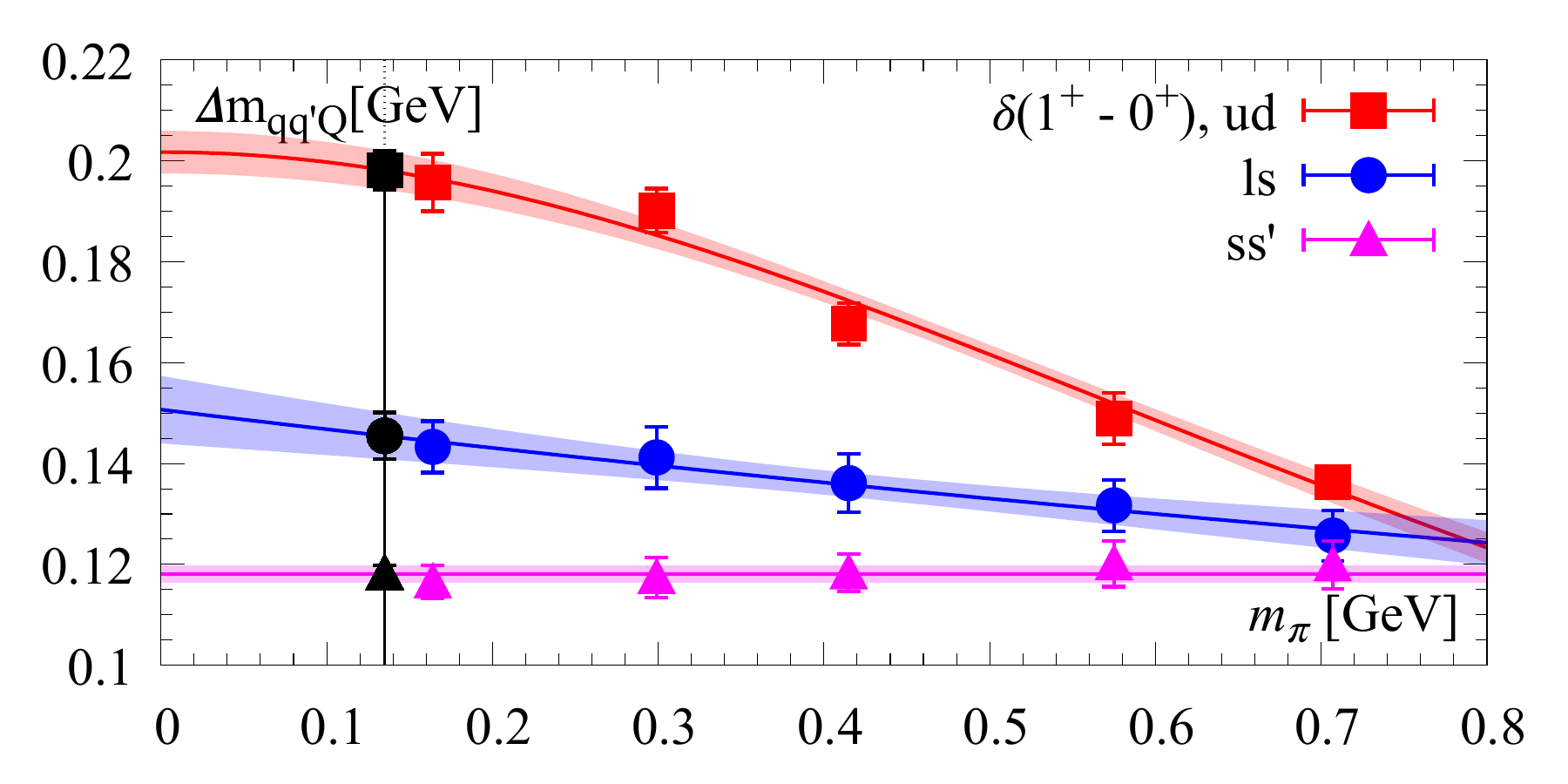}
\includegraphics[width=0.46\textwidth]{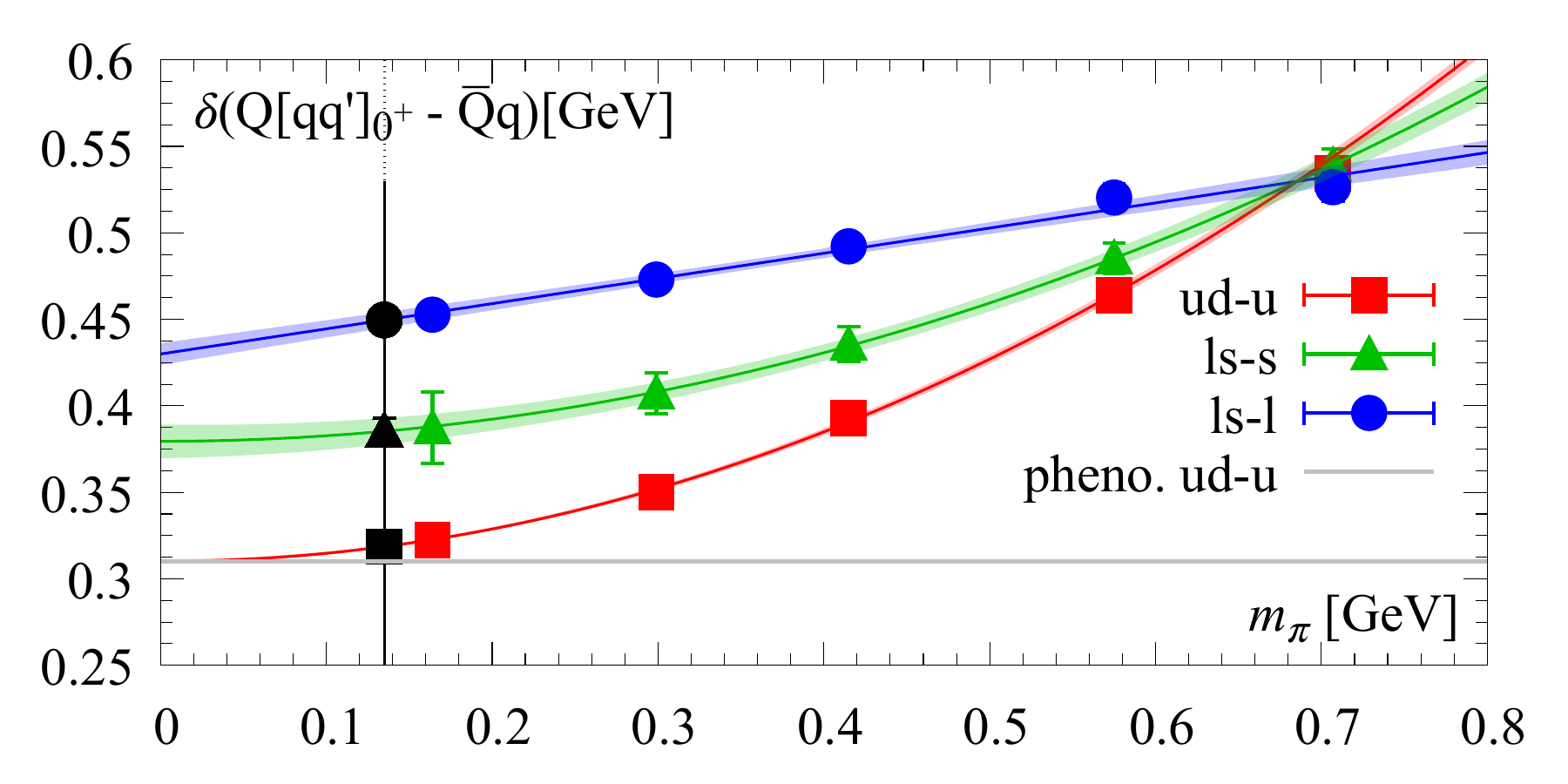}
\includegraphics[width=0.46\textwidth]{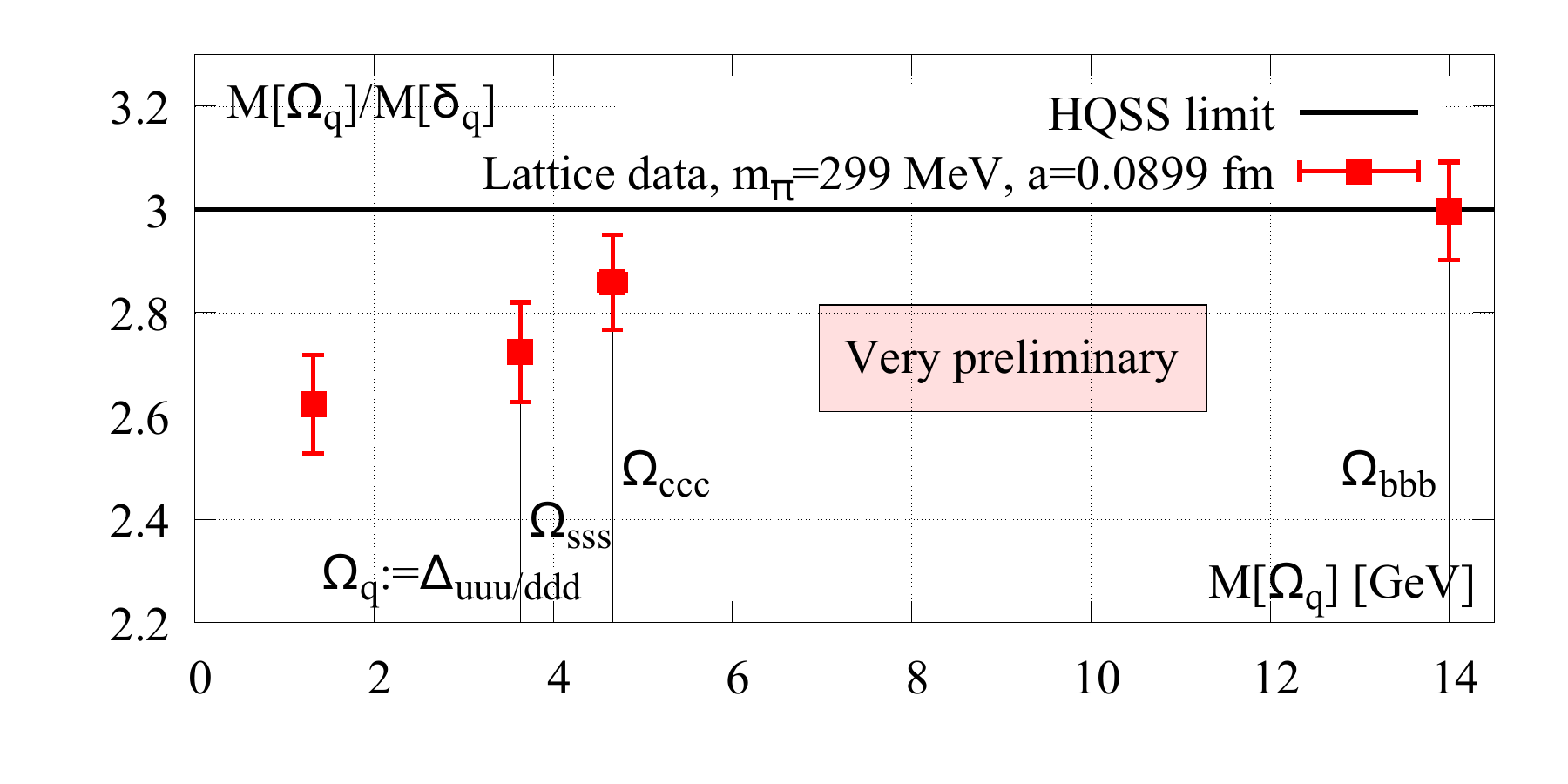}
\caption{Top left: Bad/good, not-even-bad/good diquark mass splittings over pion mass. Top right: Bad/good diquark mass splittings for different light quark flavor combinations. Bottom left: Good-diquark/quark splitting. Bottom right: Preliminary results for ratio $M_{\Omega_q}/M_{\delta_q}$. See text for details.}
\label{fig:dmass1}
\end{figure}

\section{Results}

The mass differences and density correlations are studied on the publicly available $n_f=2+1$ full QCD gauge field ensembles provided by the PACS-CS collaboration \cite{Aoki:2008sm}. The ensembles have lattice volumes of $32^3\times 64$ with a lattice spacing of $a=0.090$fm (redetermined in \cite{Francis:2016hui}). In total five ensembles with the pion masses ${m_\pi=164, 299, 415, 575}$ and ${707\,{\rm MeV}}$ are used. In all cases the strange quark mass is held fixed $m_s\simeq m_s^{\rm{phys}}$. 
For the charm quarks we use the Tsukuba-type effective relativistic heavy quark action in the tuning of \cite{Namekawa:2013vu} and the bottom quarks were handled using NRQCD, see e.g. \cite{Lewis:2008fu}. In both cases we reused the propagators gathered and presented in \cite{Francis:2016hui,Francis:2018jyb}. For the static quarks we use several types of smearing to improve their signal, see \cite{Francis:2021vrr}.\\
To extract the mass differences from the correlator data,
we follow a procedure where we determine our results from different smearing settings to analyze the correlator ratios and to then cross-check them with differences formed from analyzing the individual correlators separately. 
Our findings are shown in Fig.~\ref{fig:dmass1}: The top left figure shows the diquark-diquark splittings that were discussed above for all values of the light quark mass. The physical light quark mass is given as a vertical black line throughout. In all cases we observe a positive splitting, which indicates the good diquark is indeed lighter than all other channels. Furthermore, for the $\delta(1^+ - 0^+)$ splitting, i.e. the bad/good splitting, we observe a clear trend towards larger values as the quark mass becomes lighter. This is in-line with the expectations of the good diquark attraction picture. The not-even-bad/good splittings were extrapolated using constant fits while the bad/good splitting was extrapolated using the Ansatz: $\delta (1^+-~0^+)_{qq'} = A/\left[ 1+\big(m_\pi/B\big)^{n\in 0,1,2}\right]$, see \cite{Francis:2021vrr}.
In the right top panel of the same figure we show the bad/good splitting over pion mass but this time with varying the quark flavor content $qq'=ud,us,ss'$, where in the last case we assume two different quark flavors at the strange quark mass. The fit performs well and in-line with the good diquark attraction picture we observe that the $qq'=ud$ channel has the largest splitting.
Finally in the bottom left of the figure we show the good-diquark/quark splitting for the different possible combinations of light quark flavors. The fit Ansatz needs to be modified in this case and reads $\delta(Q [q_1 q_2]_{0^+} - \bar{Q} q_2) = C\,\left[ 1 + (m_\pi/D)^{n \in 0,1,2}\right]$. A good description of the data is achieved. 
The reader is referred to our publication \cite{Francis:2021vrr} for details and numerical values of the fits.\\
In Fig.~\ref{fig:dmass1} (bottom right) we present our preliminary, first, results on testing HQSS via the ratio ${M_{\Omega_{q}}}/{ M_{\delta_{q}} }$. Here we focus on just one ensemble with pion mass $m_\pi=299~$MeV and study the channels with $q=u,s,c$ and also $q=b$. The first three can be analyzed using the same methods as before. However, since we are using NRQCD bottom quarks in this case we study the momentum-dispersion relation to determine the kinetic mass instead. The results are plotted over the $\Omega_q$ baryon mass. We find a trend towards the HQSS limit with increasing quark mass. While there are still some deviations for the charm quark they are still within one standard deviation. Making a statement about how well HQSS is already fulfilled is premature and we hope to increase precision in the future.

\begin{figure}
\centering
\includegraphics[width=0.24\textwidth]{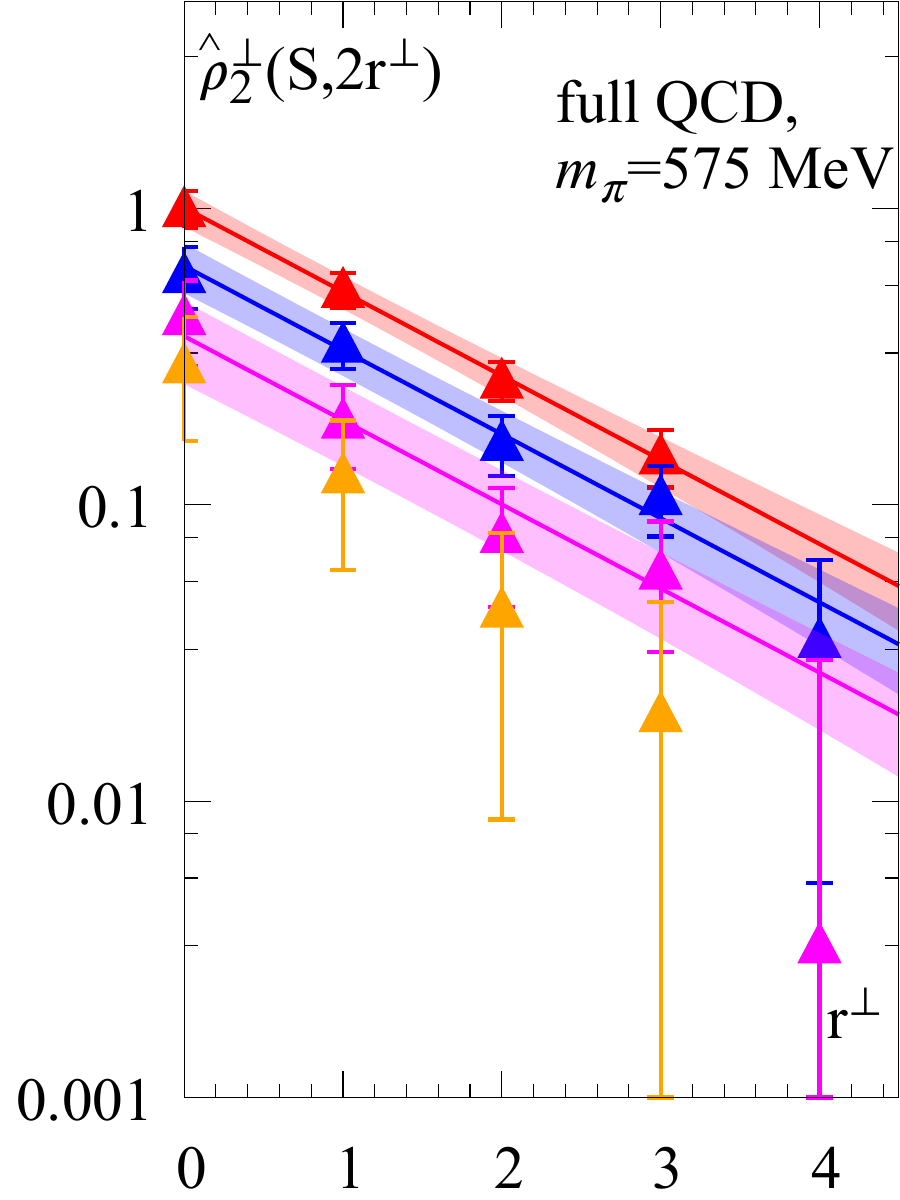}
\includegraphics[width=0.24\textwidth]{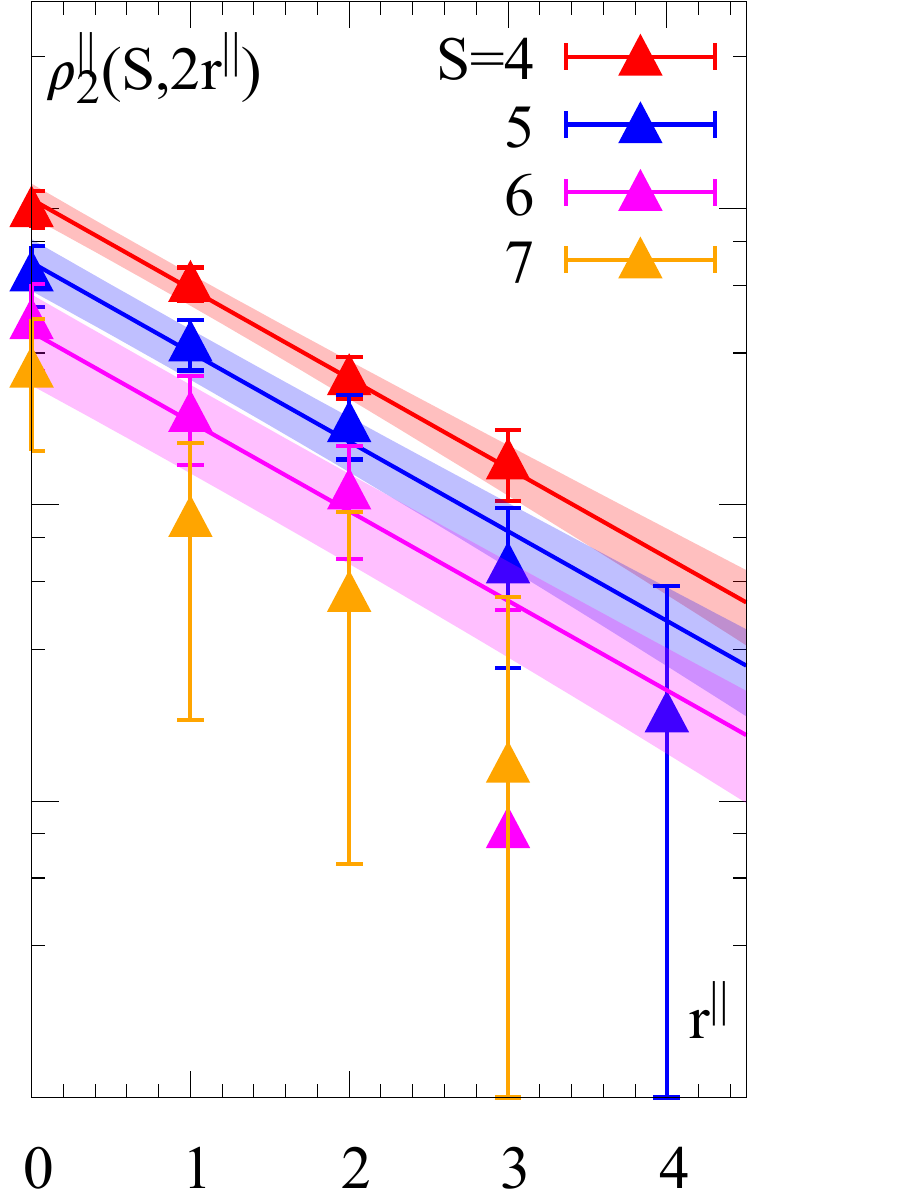}
\includegraphics[width=0.24\textwidth]{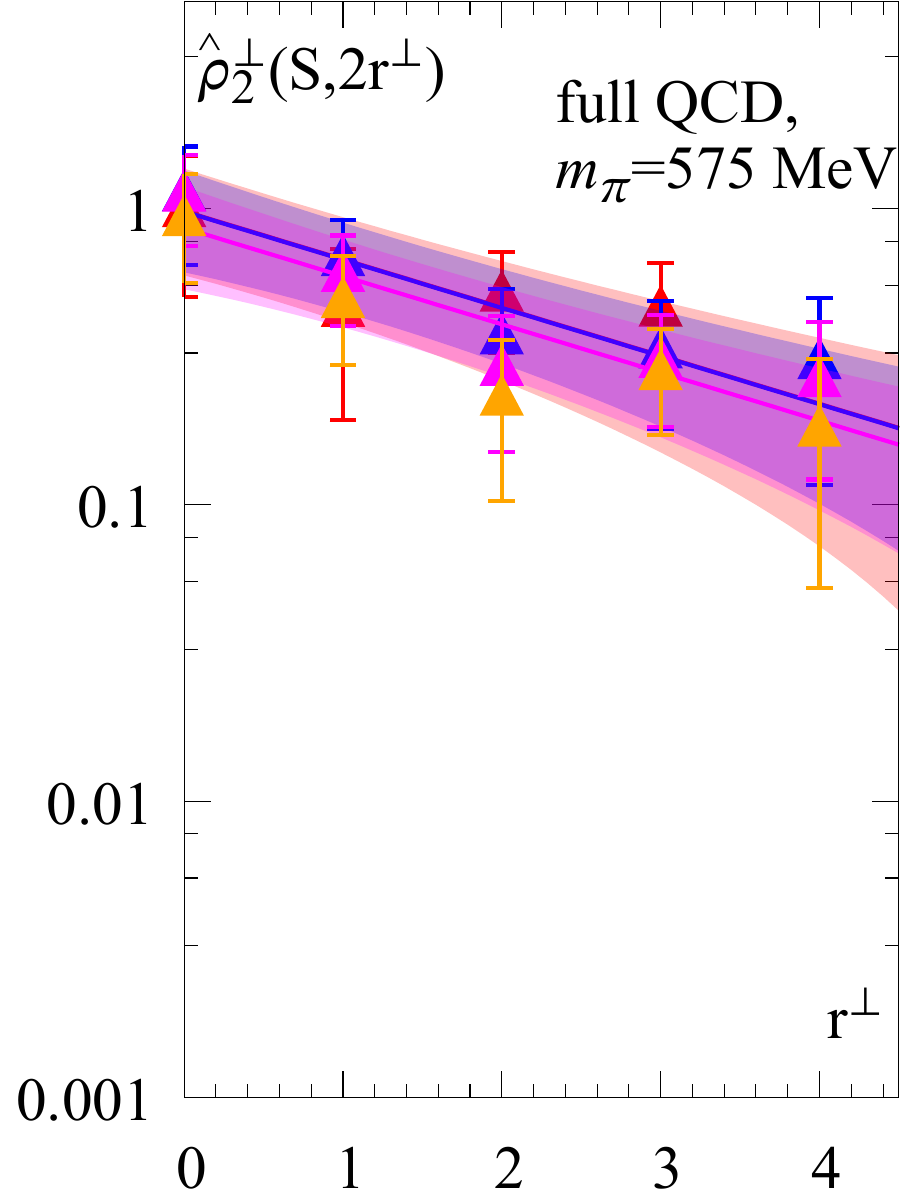}
\includegraphics[width=0.24\textwidth]{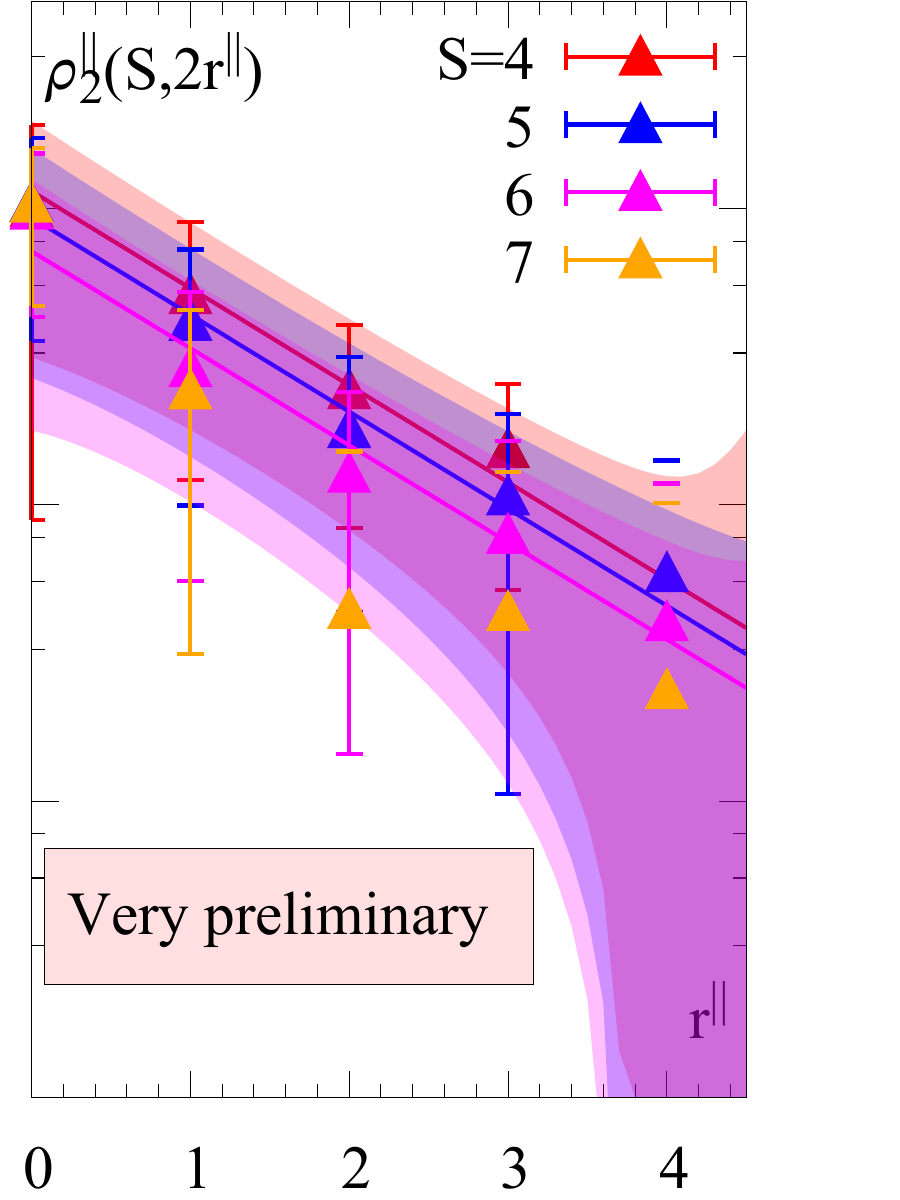}
\caption{Left: Radial and tangential spatial correlations in the $udQ$ channel. We observe the same slopes and therefore a spherical shape. Right: The same for $udQ\rightarrow uds$. We observe different slopes and a possible distortion of the correlations through the spectator quark.  }
\label{fig:dens1}
\end{figure}

After the splittings our next goal is to study the shape of the good diquarks. Results in the case where $|\vec x_1|=|\vec x_2|$, where we do not distinguish the radial and tangential correlations, strongly indicated that the only channel that exhibits an exponential decay is the good diquark channel, please see \cite{Francis:2021vrr} for details.
With the quark-quark attraction established in this channel we turn to studying the radial and tangential correlation properties and show their results in Fig.~\ref{fig:dens1}. In the left two panels we show the case for the $udQ$ channel at $m_\pi=575~$MeV. Here the spectator is static and we do not expect to observe a polarization effect. Indeed, firstly we observe the exponential decay indicative of a correlation, and therefore attraction, and secondly we see that the slopes are roughly the same even by eye. As such $r_0^\perp/r_0^\parallel \simeq 1$ within uncertainties. For a complete analysis including quark mass dependence see \cite{Francis:2021vrr}.
In the two right panels we next show preliminary results on the same ensemble, however, this time the static quark is replaced with a strange quark $udQ\rightarrow uds$. Comparing the two figures we now observe a difference in the two slopes, indicating $r_0^\perp/r_0^\parallel \neq 1$ and therefore a distorting effect. 
In future work we plan to quantify this further.

\section{Conclusions}

Using a gauge-invariant approach we studied diquark properties in $n_f=2+1$ lattice QCD. While our lattice setup lets us perform our calculations close to the physical quark mass with just a short chiral extrapolation required, the continuum limit has not been addressed.
Studying mass splittings we confirmed the special status of the good diquark configuration in the sense that it is the lightest of all possibilities and that the splitting between the bad/good diquark grows going to the physical quark mass (see \cite{Francis:2021vrr} for values). 
In general good agreement with phenomenological expectations is observed.
Going beyond our previous work we further presented our first calculation on $M_{\Omega_q}/M_{\delta_q}$, which we motivated as a test of HQSS. At this time the precision of the findings is not high enough to make statements, however, a broad trend following the expectations is observed. In the future we will perform more calculations to increase statistics and also add more gauge ensembles in order to boost precision and accuracy.
Beyond diquark spectroscopy we observe the suggested quark-quark attraction in the good diquark configuration through studying spatial quark-quark correlations. In the case of a static spectator quark the diquark further shows no preferential correlation direction, implying a spherical shape. Swapping the static spectator for a strange quark and repeating the calculation we observe a distortion of this property, indicating possible polarization and interaction. The current level of precision unfortunately is not high enough to draw firm conclusion. In the future we will increase statistics to boost precision and choose heavier spectator quark masses to track this effect in the hope of quantifying it. Further work is also necessary in improving the formalism to deal more thoroughly with the de-localization properties of the non-static propagators in this observable.

\section*{Acknowledgments}
This work is supported by the National Science and Technology Council (NSTC) of Taiwan under grant 113-2112-M-A49-018-MY3. Calculations were performed on the HPC clusters HPC-QCD@CERN and Niagara@SciNet supported by Compute Canada.

\bibliographystyle{JHEP}

\bibliography{references}


\end{document}